\documentclass[10pt, conference, letterpaper]{IEEEtran}
\IEEEoverridecommandlockouts
\usepackage{mathptmx}
\usepackage[T1]{fontenc}
\usepackage{cite}
\usepackage{amsmath,amssymb,amsfonts}
\usepackage{graphicx}
\usepackage{textcomp}
\usepackage{xcolor}
\usepackage{algorithm}
\usepackage{algpseudocode}
\usepackage{amsmath}
\usepackage{colortbl}
\usepackage{booktabs}
\usepackage{tabularx}
\usepackage{adjustbox}
\usepackage{tabularx} 
\usepackage{array} 
\def\BibTeX{{\rm B\kern-.05em{\sc i\kern-.025em b}\kern-.08em
    T\kern-.1667em\lower.7ex\hbox{E}\kern-.125emX}}
\usepackage{url}
\usepackage{subfig}
\usepackage{diagbox}
\usepackage{multirow}
\usepackage{multicol}
\usepackage{enumitem}
\usepackage[T1]{fontenc}
\usepackage{textcomp}
\usepackage{makecell}

\begin{document}
\title{Adapting Network Information into Semantics for Generalizable and Plug-and-Play Multi-Scenario Network Diagnosis}

\author{
Tiao Tan $^{1}$,Fengxiao Tang$^{1*}$,Linfeng Luo$^{1}$, Xiaonan Wang$^{2}$, Zaijing Li$^{3}$, Ming Zhao$^{1}$
\\
    $^1$Central South University \\
       $^2$Xinjiang University \\
        $^3$Harbin Institute of Technology, Shenzhen \\
    234701046, tangfengxiao, luolinfeng@csu.edu.cn,\\
       107552304984@stu.xju.edu.cn, lzj14011@gmail.com, meanzhao@csu.edu.cn
}

\maketitle

\begin{abstract}

Leverage large language model (LLM) to refer the fault is considered to be a potential solution for intelligent  network fault diagnosis. However, how to represent network information in a paradigm that can be understood by LLMs has always been a core issue that has puzzled scholars in the field of network intelligence. To address this issue, we propose LLM-based Network Semantic Generation (LNSG) algorithm, which integrates semanticization and symbolization methods to uniformly describe the entire multi-modal network information. Based on the LNSG and LLMs, we present NetSemantic, a plug-and-play, data-independent, network information semantic fault diagnosis framework. It enables rapid adaptation to various network environments and provides efficient fault diagnosis capabilities. Experimental results demonstrate that NetSemantic excels in network fault diagnosis across various complex scenarios in a zero-shot manner.

\end{abstract}



\section{Introduction}

With the growing complexity and dynamic nature of networks, Artificial Intelligence for IT Operations (AIOps), especially in critical areas like network fault diagnosis, has become increasingly challenging. Modern network AIOps rely on various specialized intelligent models~\cite{li2020research,notaro2021survey}, which have high demands for training data, particularly fault and cause data. However, real-world fault data is scarce~\cite{qureshi2023toward}. At the same time, the transfer of specialized models and their adaptation to new environments incur significant operational overhead. The future of endogenous intelligent AIOps introduces new requirements, aiming to develop generalized, adaptable universal intelligent models that do not rely heavily on proprietary data. These models should be capable of cost-effective, fast migration, and training with small or even no sample data. Recently, the emergence of powerful intelligence in LLMs has provided new insights for endogenous intelligent AIOps. The LLMs' ability to understand underlying relationships enables them to rapidly exhibit strong logical reasoning capabilities that extend beyond a single domain, which has led to remarkable achievements in fields such as medicine~\cite{tu2024towards} and law~\cite{zhou2024lawgpt}.

The current mainstream LLMs primarily rely on a single modality, while some multimodal LLMs are heavily dependent on paired multimodal data. However, the network fault data available today often exists in an unmatched multimodal form, with time-series data being the predominant modality. This makes it challenging to directly apply existing LLMs to the complex field of network fault diagnosis.

Two categories of solutions have been proposed in current research to address this issue. The first category of solutions involves aligning text and time-series data, represented by the studies of~\cite{wang2024chattime, hu2025context, jin2023time}. This category includes solutions that utilize joint embedding of time-series data and textual information. However, the primary issue with these solutions is that LLMs lack the capability to understand and align the embedded time-series data with text, thereby struggling to establish causal relationships between faults and time-series data. The underlying reason for this issue is that the data embedding only enables superficial matching and shallow correlation learning between time-series and text, lacking deeper analysis of the correlation between time-series patterns and textual semantics. Moreover, these solutions require manual configuration of time-series embedding for data preprocessing, heavily relying on expert knowledge, which renders them impractical for scenarios lacking predefined fault logics.

The second category of solutions involves employing specialized small models to process time-series data~\cite{tang2024semi}, bridging with LLMs through an intermediate layer, as represented by the studies of~\cite{tang2024large, wu2024netllm, chen2024automatic, jia2024gpt4mts}. These solutions have demonstrated effective fault diagnosis in specific scenarios due to the adaptability of small models. However, these solutions still fundamentally rely on specialized small models to interpret time-series features and do not resolve the dependency on tailored fault data. Additionally, mature models are challenging to transfer to new environments.

Therefore, to address the key technical challenges faced by LLMs in the domain of generalized fault diagnosis, this paper proposes a universal intelligent semanticization approach for fault data based on LLM. 
This method involves establishing an LLM-based adaptive semantic generation (LNSG) model to transform multimodal network states and fault data into easily comprehensible natural language descriptions. Additionally, it introduces a symbolic representations method using symbol encoding technology to effectively describe and manage complex relationships (e.g., network topology and network state machine) within network data. Through the integration of intelligent semantic and symbolic representation techniques, a LLM accessible input is constructed for further LLM-driven end-to-end network fault diagnosis.

To process the end-to-end network fault diagnosis, we further combine LNSG with fundamental LLM to finalize the fault detection and root cause reasoning. However, due to lack of knowledge in professional fields, the simple deployment of LLM is hard to accurately refer the network fault. We further leverage Retrieval-augmented Generation (RAG) into the LLM-driven fault diagnosis and design a NetSemantic framework. Through a constructed dynamic network knowledge graph (NKG) and RAG,  complex network diagnostic problems are decomposed into smaller sub-problems and resolved through phased planning. Comparing with existing fault diagnosis methods, the NetSemantic boasts robust generalization capabilities, plug-and-play functionality, and non-data-dependent characteristics. 

The main contributions of this paper are summarized as follows:

\begin{itemize}
\item We are the first to propose an automated semantic representation method for network information and faults. A LLM-based adaptive template generation algorithm is present to unify multimodal network data into a textual modality.
\item  We propose an innovative symbolic representation method to represent specific network data with complex logical relationships such as network topology, network state machine and complex device information. 

\item We built an adaptive network knowledge graph and designed a NKG-based RAG method for network reasoning.
\item  We design a LLM-driven end-to-end network fault diagnosis framework with strong generalization ability, plug-and-play functionality, and non-data-dependent characteristics, enabling zero-shot fault diagnosis. The results demonstrate its superior network diagnostic capabilities across different scenarios.

\end{itemize}

\begin{figure*}[htb]
\centering
\includegraphics[width=1\textwidth]{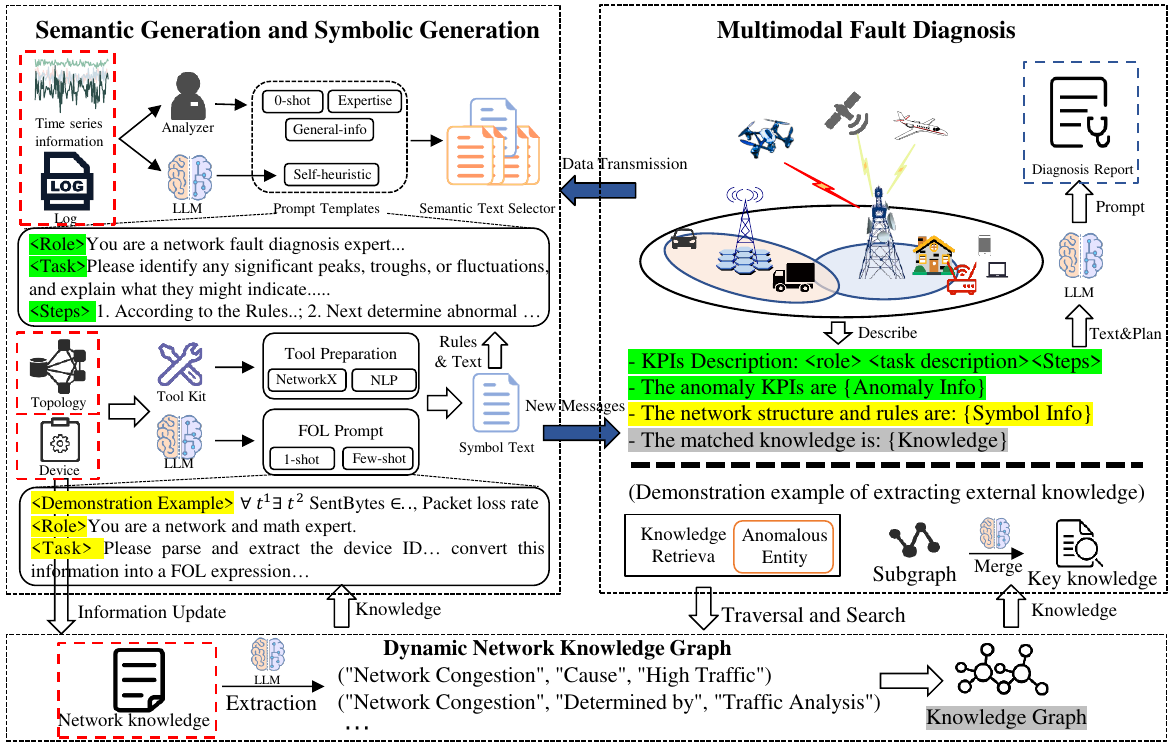}
\caption{Overview of The Workflow of Our Proposed NetSemantic Diagnostic Framework.}
\label{fig111} 
\end{figure*}

\section{Related work}
\subsection{Machine/Deep Learning Technology in Fault Diagnosis}
In recent years, network fault diagnosis has become a critical aspect of AIOps, attracting increasing attention~\cite{yedavalli2011application,lee2006network}. 
Machine learning and deep learning techniques have been widely applied to network fault detection.  Machine learning methods, by analyzing large volumes of historical fault data, can automatically uncover hidden fault patterns. In~\cite{tariq2019detecting,ren2019time,lei2020applications}, data-driven anomaly detection algorithms are proposed, which successfully identify anomalies in time-series data. However, the aforementioned methods heavily rely on data. Deep learning, especially effective with large-scale, high-dimensional data, enhances fault detection accuracy by extracting deeper features. In ~\cite{wardat2021deeplocalize, jeong2023anomalybert}, a single-model architecture is adopted, where identify anomalies is performed by introducing a supervised approach. Many other researchers have employed hybrid model architectures to fault localization~\cite{tang2024semi, cao2024advanced}, achieving efficient diagnosis by processing labeled fault datasets.

Despite these significant advancements, current techniques still face numerous challenges, particularly with imbalanced training data, high data dependency, poor model generalization, and limited model interpretability. These challenges primarily arise from the fact that faults in real-world environments often lead to severe difficulties in data collection. Furthermore, highly dynamic environments (e.g., those involving equipment upgrades or the introduction of new devices) necessitate periodic retraining of existing models to adapt to the ever-evolving network landscape.

\subsection{Application of LLMs in AIOps}
LLMs, as a significant breakthrough in the field of artificial intelligence in recent years, have achieved remarkable results in networking. Many researchers have explored the integration of specialized models with LLMs in their studies. For instance, in~\cite{tang2024large}, an LLM-assisted framework is proposed that trains an unsupervised multi-scale data anomaly detection model and incorporates multi-modal data processing driven by semantic rule trees. This framework enables LLMs to perform reasoning and generate reports. However, the method still exhibits limitations in generalizing to few-shot fault types. Another innovative work, RACopilot~\cite{chen2024automatic}, integrates LLMs in an event-driven workflow for automated diagnostics and efficient multi-source data collection, leveraging semantic understanding to predict root cause categories and generate interpretable reports. In researches on integrating specialized models with LLMs, many scholars have explored these approaches.

Moreover, to tackle the challenge of multi-modal data fusion in AIOps, NetLLM~\cite{wu2024netllm}  introduces three innovative components: a multi-modal encoder, a networked head, and data-driven low-rank adaptation (DD-LRNA). In combination with LLMs, it forms a classification system that encompasses various models, while dynamic benchmark evaluations reveal performance bottlenecks in cross-modal reasoning and hallucination suppression.

It is worth noting that recent studies have begun to focus on directly leveraging the generalization capabilities of LLMs to understand the semantics of time-series data~\cite{xue2023promptcast,liu2024can}. For instance, ChatTime, introduced in~\cite{wang2024chattime}, innovatively treats time-series as a "foreign language". By extending the vocabulary and fine-tuning pre-trained language models, ChatTime successfully achieves a unified understanding and generation across both time series and textual modalities. In~\cite{hu2025context}, researchers argue that the ability of LLMs to process language stems more from their deep understanding of linguistic logic and structure rather than merely aligning surface lexical patterns. As a result, they proposed a dual-scale GNN-based context alignment framework (DSCA-GNN) that enhances LLMs' semantic understanding of time series data through structural logic alignment. 

Despite the powerful capabilities of LLMs in processing linguistic information, the applications of them in network fault diagnosis still faces several challenges. Current methods predominantly rely on small models for network tasks, leading to a heavy dependence on data and hindering mature models from achieving robust generalization performance. In response to these challenges, this paper proposes a LLM-driven network fault diagnosis framework designed to operate efficiently across diverse network environments and overcome the limitations of existing methods in terms of data dependency and adaptability.

\section{NetSemantic Fault Diagnosis Framework}
In this paper, we propose a general end-to-end network fault diagnosis method named NetSemantic. The architecture of NetSemantic, as illustrated in Figure ~\ref{fig111}, consists of three main stages: semantic and symbolic generation, NKG construction, and fault diagnosis and report generation. 

In the first Semantic Generation and Symbolic Generation stage, we propose a LLM-based network semantic generation (LNSG) model to transform multimodal network states and fault data into a "language" (e.g. semantic and symbolic) task, enabling efficient context-aware detection and diagnosis in data-constrained network environments. Then a dynamic NKG is constructed to assist the LLM-driven reasoning process. Finally, the multimodal fault diagnosis consists of three sub-stages of processing, planning, and reasoning. In the following sections, we provide a detailed explanation of each module.

\begin{figure}[t]
\centering
\includegraphics[width=0.9\columnwidth]{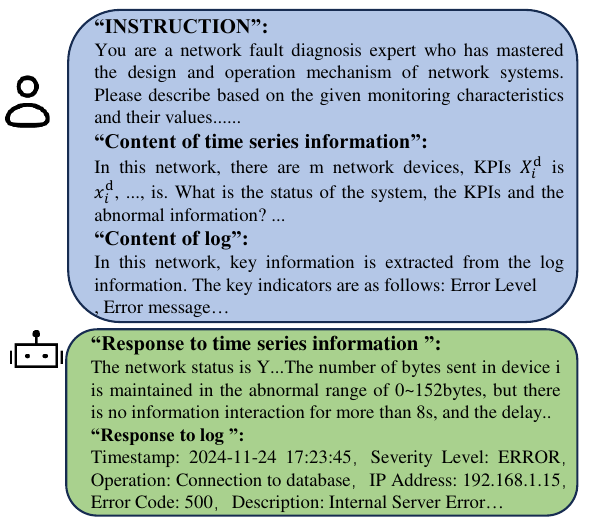}
\caption{Mapping Rules of Network Data Time Series.}
\label{fig2} 
\end{figure}

\textbf{(1) Semantic Generation and Symbolic Generation.} To establish a semantic description of real network fault data, we adopt widely used strategies and design a novel adaptive prompt template. This template utilizes a LLM to convert static and dynamic data from network faults into natural language descriptions. The template is based on real-time information such as device details and historical data in the current network environment, and it updates the template information through a knowledge graph. 

We then introduce a symbolic representation to further improve the accuracy and conciseness of semantic representations. We use several examples from a first-order logic training dataset as templates to prompt the LLM for network symbolic generation. These templates can be classified into one-shot (with a single demonstration) or few-shot (with multiple demonstrations) methods. Finally, the collected semantic and symbolic descriptions are used as input for the diagnostic module.

\textbf{(2) Dynamic Network Knowledge Graph.} The dynamic NKG is designed to provide NetSemantic with essential external network fault and device-related knowledge. It includes the following ports: (i) Construction: Leveraging the LLM, we extract network knowledge in the form of triples to construct the NKG. (ii) Updating: The NKG is online updated with real-time network data during each diagnosis process. (iii) Retrieval: During fault diagnosis, RAG queries abnormal entities from the NKG, and relevant network knowledge is extracted to support diagnosis.

\textbf{(3) Multimodal Fault Diagnosis.} With the support of the LLM, the NetSemantic system performs network fault diagnosis by decomposing the network fault diagnosis problem into smaller sub-stages of preprocessing, planning, and reasoning. 
In the preprocessing stages, the LLM concatenates the generated semantic and symbolic tokens with premise and problem statements. Then, a detailed diagnostic blueprint is automatically generated by processing the RAG from the dynamic NKG, covering all possible faults that exist. Finally, the system derives the diagnosis report through step-by-step reasoning following the blueprint.

\section{Methodology}
\subsection{LLM-based Network Semantic Generation Algorithm}


In this section, we propose a LLM-based network semantic generation model, which transforms network data collected during device operation into semantic and symbolic representations. The model is primarily divided into two components: semanticization and symbolization.

\subsubsection{Semanticization} the semanticization process comprises two components: a self-heuristic network semantic prompt and a semantic text selector.

Although carefully designed prompt templates perform well in simple tasks, they often result in unstable text quality and inconsistent diagnostic outcomes when applied to tasks that require complex logic and domain-specific knowledge. These limitations restrict their applicability across different network scenarios. To address these challenges, we propose a self-heuristic prompt template. This approach significantly enhances the semantic generation capability and stability of LLMs in network fault diagnosis.

\textbf{Self-heuristic network semantic prompts:} in networks, the time series is generally the primary representations of dynamic network data, which is typically stored in chronological order, with each row representing a data record associated with a specific timestamp. However, time series is hard for LLMs to directly understand and reason about~\cite{wang2024chattime}. 
To assist LLMs in better understanding these dynamic network data and improve the accuracy of subsequent diagnostic results, we present a set of mapping rules between time series of dynamic network data and semantic text, structuring data from $m$ devices and $n$ key performance indicators (KPIs) into the form shown in Figure ~\ref{fig2}. In this structure, $X_{\mathrm{i}}^{\mathrm{d}}$ and $x_{\mathrm{i}}^{\mathrm{d}}$ represent the feature variable names and their corresponding values for the $i$-th KPI of the $d$-th device, respectively.

\begin{table}[t]
\caption{Semantic Prompt Generation}
\centering
\resizebox{1\columnwidth}{!}{
\begin{tabular}{lc}
\toprule[1.2pt]
Template Name & Template \\
\midrule
\midrule
self-heuristic & \makecell[{{l}}]{\textbf{INSTRUCTIONS} <role> <task description> <reinforce>\\
\textbf{CONTENT} <input> <task confirmation>\\
\textbf{CONSTRAINT} <knowledge><steps><rules>} \\
\bottomrule[1.2pt]
\end{tabular}
}
\label{tab0}
\end{table}

\begin{table}[t]
\caption{Symbolic Prompt Generation}
\centering
\resizebox{1\columnwidth}{!}{
\begin{tabular}{lc}
\toprule[1.2pt]
Template Name & Template \\
\midrule
\midrule
one-shot & \makecell[l]{\textbf{CONTENT} <task description><demonstration example>\\<input>} \\
\midrule
few-shot & \makecell[l]{\textbf{CONTENT} <task description><demonstration example> \\ \dots <demonstration example><input>} \\
\bottomrule[1.2pt]
\end{tabular}
}
\label{tab111} 
\end{table}

In Table~\ref{tab0}, we provide a detailed explanation of the method used to construct the self-heuristic prompt templates. The template mainly consists of instructions, content, and constraints. The part of instructions includes <role> (which requires the model to assume a role related to the task), <task description> (providing contextual information before presenting each device), and <reinforce> (emphasizing key points that need to be reiterated). The part of content includes <input> (network data) and <task confirmation>. The part of constraints includes <knowledge> (external network knowledge), <steps> (the model thinks step by step), and <rules> (semantic generation rules). 

It is important to note that the content of the template does not require manual design. Instead, it is automatically updated by extracting <role><task description><knowledge> from the NKG based on the current device information and historical data from the network. This ensures that the template is adaptable to different network scenarios.

Due to factors such as the inherent random sampling mechanism of LLMs, contextual variations, model updates, and potential hardware or software differences, even when using the same network semantic prompts, the generated semantic outputs often exhibit discrepancies. These differences can lead to omissions or redundancies in the semantic descriptions. To address this issue, we propose the semantic text selector, which is an innovative approach that efficiently extracts the optimal semantic description.

\textbf{Semantic text selector:} first, we perform $n$ semantic sampling iterations on different network semantic prompts using LLM to obtain a set of semantic texts $T=\{t_1,t_2,...,t_n\}$.
To conduct a quantitative analysis of these texts, we use the pre-trained Sentence-BERT~\cite{reimers2019sentence} model to encode the texts, resulting in normalized semantic vectors $v_i$, as follows:

\begin{equation}  
\mathbf{v}_i=f_{\mathrm{SBERT}}(t_i)\in\mathbb{R}^d,\quad\|\mathbf{v}_i\|_2=1,
\end{equation}
where $f_{\mathrm{SBERT}}$ represents the encoding function of Sentence-BERT, and $d$ is the embedding dimension. After obtaining the normalized semantic vectors $\mathbf{v}_i$, we further construct a fully connected cosine similarity matrix $S\in\mathbb{R}^{n\times n}$, where the semantic similarity between a pair of semantic texts $(t_i,t_j)$ is given by:

\begin{equation}
S_{i,j}=\mathbf{v}_i^\top\mathbf{v}_j,
\end{equation}
to measure the semantic representativeness of each text 
$t_i$, we define its mean centrality index $\mu_i$, which can be expressed as:

\begin{equation}
\mu_i=\frac{1}{n-1}\sum_{j=1,j\neq i}^nS_{i,j},
\end{equation}
this index reflects the degree of aggregation of $t_i$ in the semantic space. Therefore, the semantic text 
$t_{\mathrm{best}}=\arg\max_{t_i\in T}\mu_i$ with the highest average similarity is selected as the optimal result.

It is important to note that this strategy not only effectively captures key information in network semantic representations but also helps reduce the token input size without compromising the completeness of the description.

\subsubsection{Symbolization}

The proposed semanticization is good at handling the time series of network data, however, data such as network topology (e.g., graph, table or chain), network protocols (e.g., unstructured document, interrelated tables and symbols), and device configuration are difficult to directly semanticize due to redundant information and complex logic. Using symbols to convey fault information provides significant advantages for those certain types of data. For instance, the network topology and device configuration data can be represented using first-order symbols to denote nodes, edges, and their connections, while network transmission rules and routing protocols can be described using second-order symbols. This symbolic approach not only reduces data redundancy but also effectively minimizes the number of tokens required for processing through standardized representations~\cite{gavrilo2024using}. Moreover, it helps mitigate issues such as misdiagnosis and overlooked faults during the inference process. Therefore, we propose an innovative symbolic representations method that leverages the generative capabilities of LLMs to design a symbolic template for transforming network information into symbolic form.


Firstly, for one-shot and few-shot prompting, as shown in Table~\ref{tab111}, the primary benefit is that it helps the LLM understand the structure of first-order symbolic outputs through demonstration examples. This allows the LLM to convert network topology information, port status, device states, and other relevant data into symbolic representations. Furthermore, the demonstration example serves as a universal template, applicable to various network environments.

Second, each KPI value is determined by multiple attributes. We construct a feature vector matrix for each KPI, where each device can form a feature vector $S$ through its set of symbolic attributes.


\begin{equation}  
\mathbf{S}=[\mathrm{KPI}_1,\mathrm{KPI}_2,\mathrm{KPI}_3,\cdots,\mathrm{KPI}_m],
\end{equation}
this feature vector represents the different KPI values for each device. When determining the normal range for KPIs, the necessary device attributes can be retrieved through the knowledge graph. The above information is then transformed into second-order symbols to capture the rule-based information within the network.

However, LLMs have inherent limitations in handling symbolic reasoning tasks, particularly in scenarios that require extensive mathematical reasoning. During the symbolic generation process, we do not focus on deriving network rules from basic parameter information. Therefore, we integrate the Z3 solver, which is based on the CDCL optimization search algorithm~\cite{de2008z3}. It can efficiently handle logical constraints, enabling the inference of network information to generate network rules, thus improving overall symbolic generation efficiency.


\begin{figure*}[htb]
\centering
\includegraphics[width=1\textwidth]{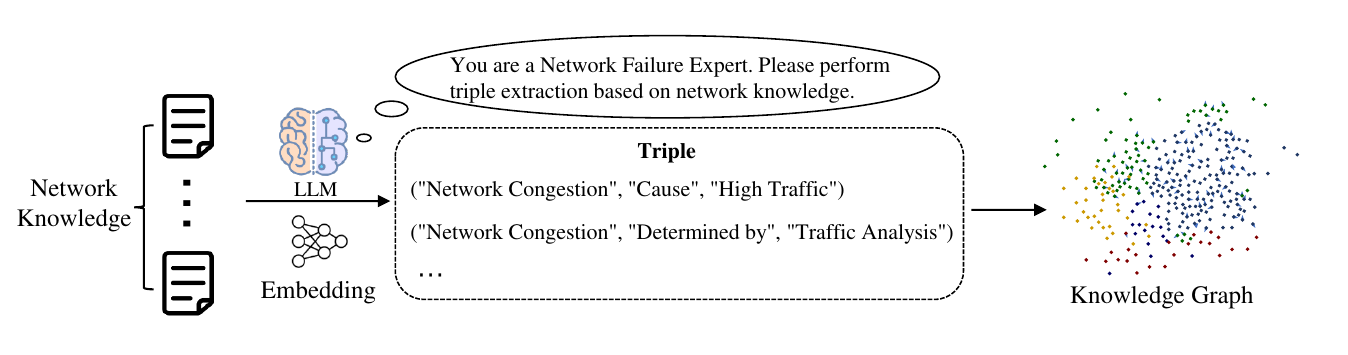}
\caption{Network Knowledge Graph Construction.}
\label{fig8}
\end{figure*}

\subsection{Network Knowledge Graph}
\subsubsection{Knowledge Graph Construction}
The NKG serves as an external knowledge base to guide the LLM in fault diagnosis. It contains data on all critical features, including information about entity devices and fault details. It is important that the knowledge within the knowledge base is rich and capable of providing accurate insights. We classify common fault types according to the seven-layer network protocol, ensuring maximum coverage of potential faults across different layers. Additionally, we collect theoretical metrics for various devices to calculate their KPI thresholds. Values exceeding these thresholds are flagged as anomalies. To streamline fault modeling and data processing, as shown in Fig.~\ref{fig8}, we use a triple-based format to represent relationships among network devices, performance metrics, and fault types, storing specific data in the following format:

\begin{equation}  
  G=\{E,R,F\},
\end{equation}
$G$ represents the network fault knowledge graph. $E$ is the set of key entity information used to identify network faults and derive rules from the knowledge base, denoted as $\{e_1,e_2,...,e_n\}$. This set includes network devices, performance metrics, fault types, state descriptions, protocols, and other relevant information. $R$ represents the set of relationships $\{r_1,r_2,...,r_n\}$, where each relationship $r$ acts as an edge in the knowledge graph, indicating connections between different entities. $F$ refers to the set of facts $\{f_1,f_2,...,f_n\}$, which describes the states, events, attributes, and specific manifestations of the entities. All knowledge extraction is based on large language models. 

Notably, the construction of the knowledge graph not only ensures the completeness of the content but also reduces token input by 40\%, enhancing efficiency.

\subsubsection{Knowledge Graph Update}
In a highly dynamic network environment, frequent upgrades of network devices and the introductions of new devices often lead to node instability, resulting in outdated information in the NKG. To address this issue, we propose an adaptive data update mechanism that dynamically incorporates current network data and diagnostic results into the knowledge graph, replacing the outdated entity information. All knowledge updates are based on LLMs.

\subsubsection{Knowledge Graph Retrieval}




We have identified several issues with traditional knowledge graph retrieval, such as insufficient semantic understanding, which often leads to the failure to retrieve the desired subgraphs. To address this, we vectorize the query content and the entities in the knowledge graph using Word2Vec~\cite{jatnika2019word2vec}. By calculating the cosine similarity between the query content and the entities in the graph, we quantify the semantic relevance. After calculating the similarity, we rank the entities based on their similarity and select the top-$K$ entities as candidate central nodes, denoted as $\{p_1, p_2, ..., p_k\}$.

To ensure comprehensive and efficient knowledge retrieval, we perform multi-hop neighborhood expansion on the candidate central node $p_i$, aiming to cover as much of its associated network information as possible.

\subsection{Fault Diagnosis and Report Generation}
Next, we will explain how to perform fault reasoning and analysis according to the planned approach, utilizing the collected semantic and symbolic descriptions.

\subsubsection{Token Concatenation}
In the preprocessing stage of network fault diagnosis, we divide the entire network description into four components: network semantic information, network symbolic information, network knowledge, and problem statements. We have obtained the network semantic and symbolic information from the Semantic Generation and Symbolic Generation modules. By reasoning over the semantic information using the network rules embedded in the symbolic information, we can identify whether there are any anomalies in the network and extract the key entities from them. Let the given anomalous information be denoted as $T$, and the set of anomalous entities as $N = \{n_1,n_2,...,n_m\}$. Based on these key entities $N$, we perform subgraph matching on the knowledge graph through semantic similarity-based retrieval in order to retrieve relevant network knowledge. Additionally, problem statements are formulated by analyzing the extracted abnormal information to describe the faults or anomalies present in the current network. Finally, we integrate the information from these four components to provide more effective support for subsequent fault diagnosis and reasoning.

\subsubsection{Automatic Blueprint Generation}
In network fault diagnosis, multiple fault causes can lead to the same observable fault phenomenon, resulting in a set of potential fault blueprints $Z=\{z_1,z_2,...,z_m\}$, where each $z_i$ represents a potential factor contributing to the anomaly. By establishing a one-to-one correspondence between the blueprint set $K=\{k_1,k_2,...,k_m\}$ and the fault cause set $Z$ (i.e., $k_1$ corresponds to $z_1$, $k_2$ to $z_2$, etc.), we can systematically diagnose the underlying causes and identify the specific fault responsible for the observed anomaly. For example, the matching results may return fault causes such as "insufficient network bandwidth" or "weak transmission signal."

This approach not only allows for the rapid identification of relevant network knowledge but also ensures comprehensive coverage of all possible fault sources, thereby improving the accuracy and efficiency of fault diagnosis.

\subsubsection{Diagnosis Report Generation}
Next, for the generated blueprint $z_i$, a corresponding diagnostic plan is developed by extracting the diagnostic steps $S_i = \{s_1,s_2,...,s_p\}$ associated with $z_i$ from the NKG. For example, bandwidth congestion may require checking traffic statistics, routing issues may necessitate inspecting the routing table, and hardware failures may require checking the status of the network interface card (NIC), etc. These diagnostic checks are then performed. In the reasoning and analysis phase, a LLM is employed to infer and synthesize multiple potential causes, ultimately providing a final fault report. This report includes the fault type, fault phenomenon, fault explanation, summary, and proposed solutions.


\section{Experiments}

\begin{table*}[t]
\caption{Performance Comparison of Different Models}
\label{1}
\centering
\begin{tabular*}{\textwidth}{@{\extracolsep{\fill}}l *{5}{c} *{5}{c} } 
\toprule[1.2pt]
\multirow{2}{*}{Model} &\multicolumn{4}{c}{Anomaly Detection} & \multicolumn{4}{c}{Fault Diagnosis} \\ 
\cline{2-5} \cline{6-9}
& Accuracy$ \uparrow$ & Recall$ \uparrow$ & FNR$ \downarrow$ & FPR$ \downarrow $&Accuracy$ \uparrow$ &Recall$ \uparrow$ &FNR$ \downarrow$ &FPR$ \downarrow$ \\
\hline
\hline
{SR-CNN}&84.34&93.74&6.26&74.31&49.74&67.19 &32.81 &43.28\\
{CL-MPPCA}&85.33&88.29&11.71&33.13&68.91&59.10&40.90&\textbf{1.44}\\
{ANOMALYBERT}&85.39&90.34&9.66&45.50&65.81& 56.81 & 43.19 & 4.41  \\
{LSTM-Transformer}&88.49&96.10&3.90&59.00&72.57&80.22&21.06&24.34\\
{FTS-LSTM}&89.10&91.51&8.49&25.95&81.15&87.72&12.28&11.63\\
\hline
{NetSemantic}&\textbf{96.10}&\textbf{98.16}&\textbf{1.84}&\textbf{14.56}&\textbf{89.49}&\textbf{89.54}&\textbf{10.46}&1.75\\
\bottomrule[1.2pt]
\end{tabular*}
\end{table*}

\begin{figure}[t]
\centering
\subfloat[]{\includegraphics[width=0.5\columnwidth]{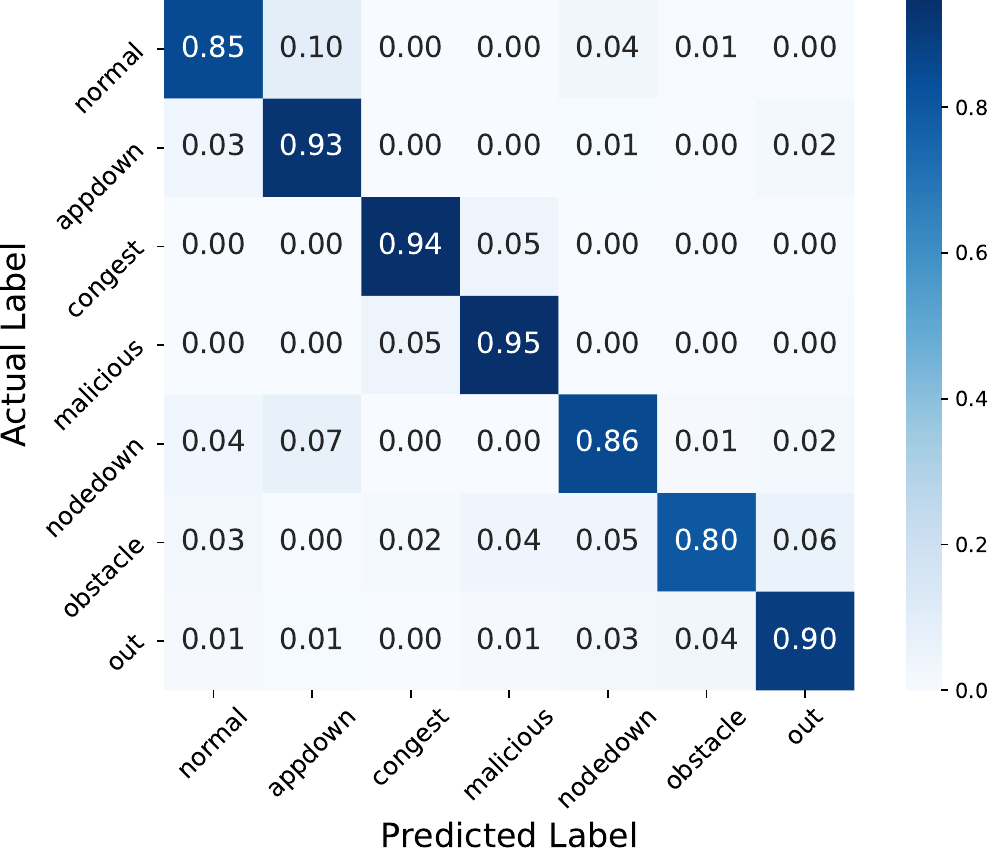}}
\hfill
\subfloat[]{\includegraphics[width=0.5\columnwidth]{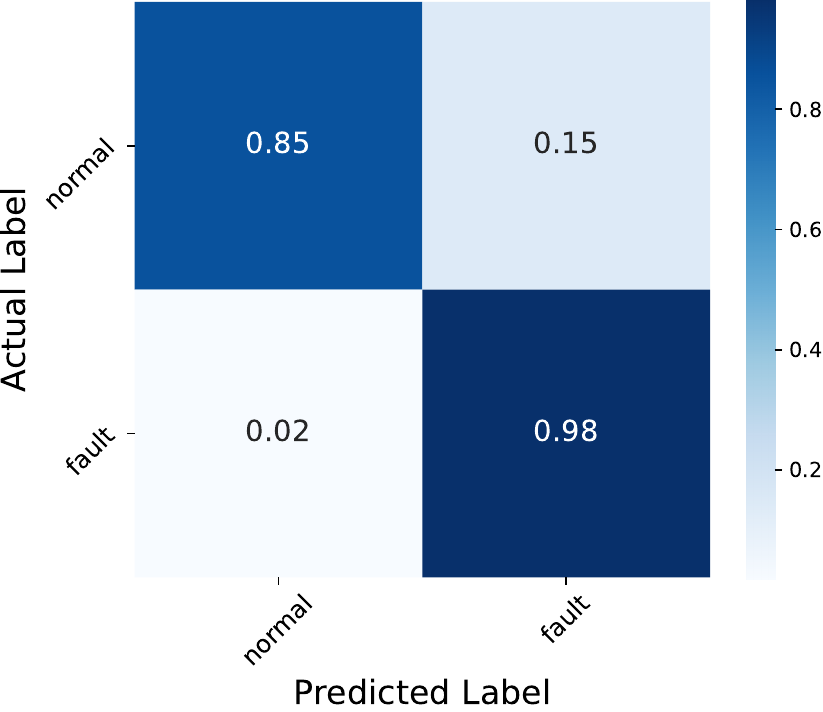}}
\caption{Confusion Matrix for NetSemantic Fault Diagnosis (a) and Anomaly Detection (b). The confusion matrix shows whether the model accurately identifies true anomalous samples and has fewer errors mislabeling correct samples as anomalous.}
\label{fig4}
\end{figure}


\begin{figure}[t]
\centering
\subfloat[]{\includegraphics[width=0.5\columnwidth]{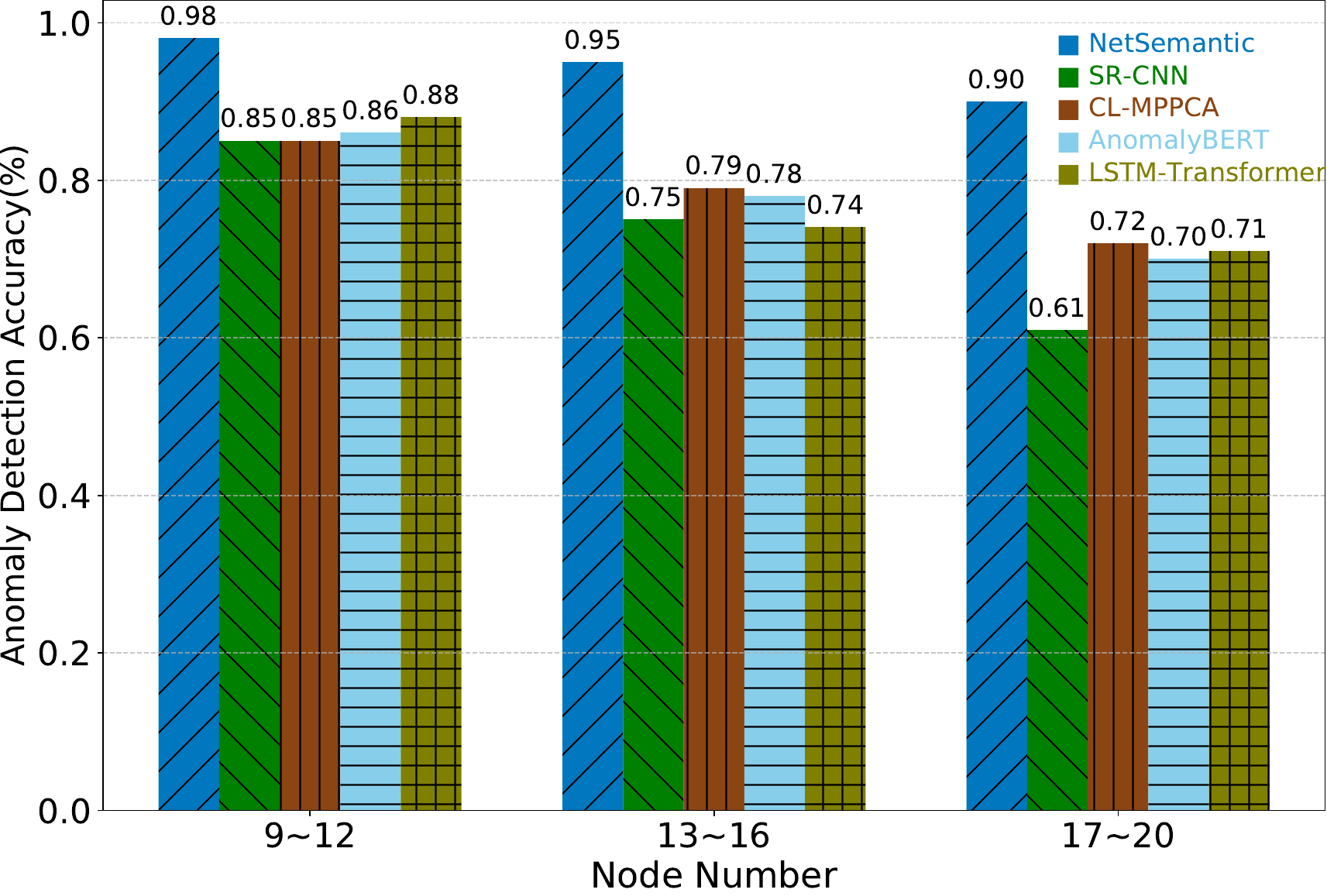}}
\hfill
\subfloat[]{\includegraphics[width=0.5\columnwidth]{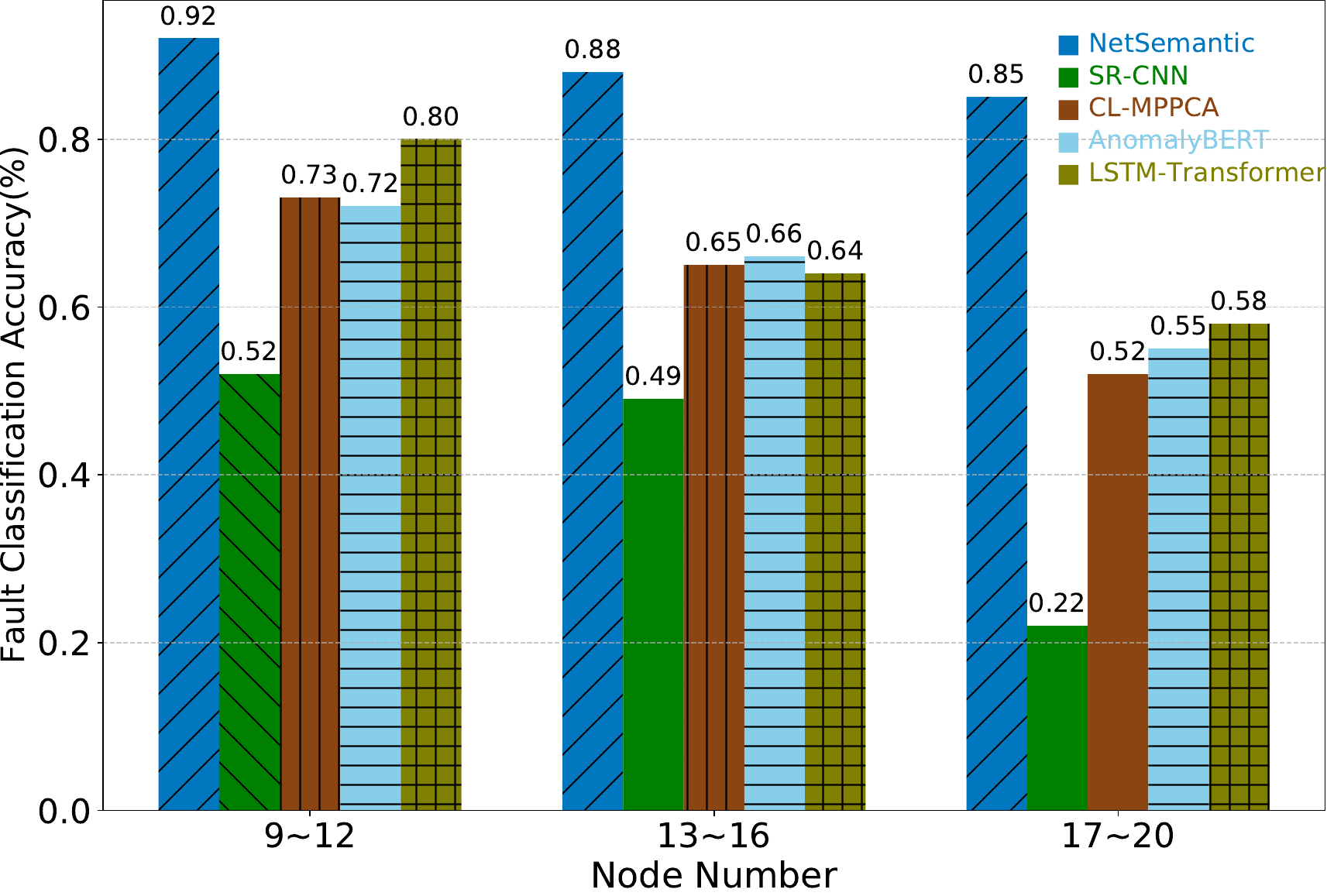}}
\caption{Anomaly Detection Accuracy (a) and Fault Diagnosis Accuracy (b) under networks of different scales. The comparison was performed on networks with node counts of 9-12, 13-16, and 17-20 nodes.}
\label{fig6}
\end{figure}

\begin{figure}[t]
\centering
\includegraphics[width=0.9\columnwidth]{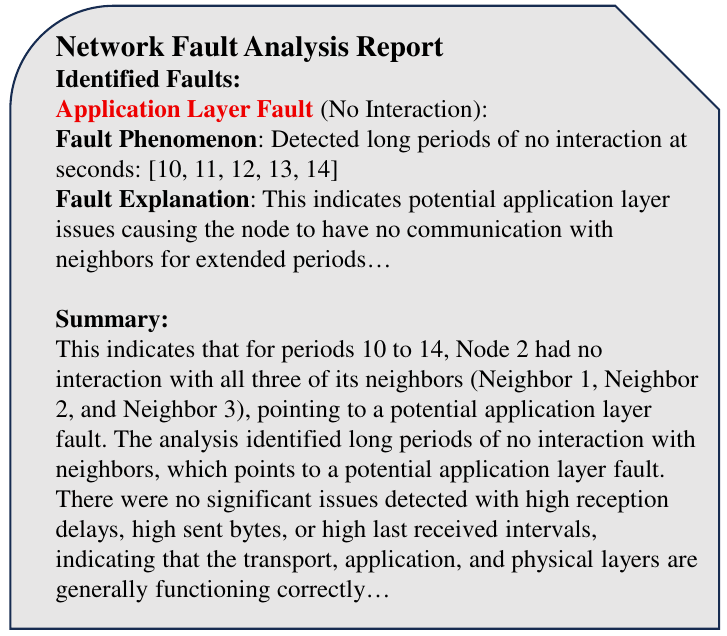}
\caption{Fault Diagnosis Report.}
\label{fig11}
\end{figure}

\subsection{Environment Setup}
{\bfseries Model.} To validate and evaluate the performance of the framework, experiments were conducted using large language models such as GPT-4o~\cite{islam2024gpt} and GPT-3.5~\cite{ouyang2022training} as the underlying models.\par

{\bfseries Dataset.} The network fault dataset classifies network failures into six major categories: 1. Application Crash; 2. Malicious Traffic; 3. Network Congestion; 4. Network Node Crash; 5. Out of Communication Range; 6. Communication Obstacles. This dataset is collected from a digital twin experiment with embeded semi physical platform and network simulator of NS-3~\cite{tang2024semi}\footnote{\url{https://github.com/SmallFlame/muti-network-data}}.\par

{\bfseries Baseline.} In this section, we compare the fault diagnosis results diagnosed using the LLM with several popular ML/DL-based baselines. \par
\begin{itemize}
\item CL-MPPCA~\cite{tariq2019detecting} employs both neural networks and probabilistic clustering to enhance anomaly detection performance.
\item SR-CNN~\cite{ren2019time} combines the Sparse Representation (SR) model with the CNN model, improving the precision of time series anomaly detection.
\item AnomalyBERT~\cite{jeong2023anomalybert} built on the Transformer architecture, it is designed to discern temporal contexts and identify unnatural sequences.
\item LSTM-Transformer~\cite{cao2024advanced} introduces a novel hybrid architecture that combines LSTM and Transformer, specifically tailored for multi-task real-time prediction.
\item FTS-LSTM~\cite{tang2024semi} incorporates an attention mechanism into the LSTM architecture to enhance the focus on critical information.

\end{itemize}

\begin{table*}[htb]
\caption{Ablation Study}
\label{2}
\centering
\begin{tabular*}{\textwidth}{@{\extracolsep{\fill}}l *{5}{c} *{5}{c} } 
\toprule[1.2pt]
\multirow{2}{*}{Model} &\multicolumn{4}{c}{Anomaly Detection} & \multicolumn{4}{c}{Fault Diagnosis} \\ 
\cline{2-5} \cline{6-9}
& Accuracy$ \uparrow$ & Recall$ \uparrow$ & FNR$ \downarrow$ & FPR$ \downarrow $&Accuracy$ \uparrow$ &Recall$ \uparrow$ &FNR$ \downarrow$ &FPR$ \downarrow$ \\
\hline
\hline
{Net}&62.43&64.61&35.39&75.90&45.45& 45.24 & 54.76 & 53.12  \\
{Net-NoKG}&76.92&78.49&21.51&67.64&56.74&56.31 &43.69 &40.16\\
{Net-GPT3.5}&83.31&83.69&16.31&35.00&70.42&70.14&29.86&27.27\\
{Net-NoSymb}&88.21&88.43&11.57&25.12&79.62&79.26&20.73&17.30\\
{NetSemantic}&\textbf{96.10}&\textbf{98.16}&\textbf{1.84}&\textbf{14.56}&\textbf{89.49}&\textbf{89.54}&\textbf{10.46}&\textbf{1.75}\\
\bottomrule[1.2pt]
\end{tabular*}
\end{table*}

\subsection{Performance Comparison}

In Table~\ref{1}, we present a detailed comparative analysis of NetSemantic and various other models based on classification accuracy, recall rate, false negative rate (FNR), and false positive rate (FPR) for anomaly detection. The metrics are defined as follows: Classification accuracy represents the proportion of fault samples correctly classified; detection precision denotes the accuracy of successfully detected fault samples; recall rate (or fault detection rate) indicates the proportion of fault samples correctly identified as such; the false negative rate (FNR) measures the proportion of fault samples incorrectly classified as normal; and the false positive rate (FPR) represents the percentage of normal samples mistakenly classified as faults. We evaluated the performance of these metrics, and the results show that NetSemantic demonstrates superior performance on several key indices, achieving a 5\% to 10\% improvement over several popular ML/DL-based baseline methods. Furthermore, compared to other methods, NetSemantic enhances the interpretability of the data and increases engineers' trust in the detection results. It also facilitates the rapid identification of potential issues, effectively reducing the risk of misjudgments, improving the efficiency of problem resolution, and ultimately enhancing the overall reliability of fault management.


Fig.~\ref{fig4} present the confusion matrix analyses of anomaly detection results generated by our NetSemantic model on networks with 9 to 20 nodes. Specifically, Fig.~\ref{fig4}(a) primarily evaluates the model’s accuracy in identifying various anomalies, highlighting its high precision in classifying most anomaly types. In contrast, Fig.~\ref{fig4}(b) illustrates the overall accuracy of the anomaly detection process, where the recognition rate of anomalous samples reaches up to 98\%, demonstrating the model's capability to analyze nearly all anomalous network information.

Then, we further examined the impact of network scale on diagnostic performance, as depicted in  Fig.~\ref{fig6}. With an increase in network size, diagnostic results are somewhat affected—mainly due to the influx of extensive multimodal network data, which can occasionally lead to the oversight of certain information and subsequent diagnostic errors. Nevertheless, in terms of anomaly detection and classification accuracy, our model significantly outperforms other existing models.

Fault Diagnosis Report: As shown in Fig.~\ref{fig11}, we present a selection of anomaly reports generated during network anomaly events. These reports primarily include descriptions of key indicators such as packet transmission rate, error rate, and latency, along with the anomalies identified by the model. The reports reveal that Node 2 experienced prolonged inactivity, which is consistent with the observed real-world scenario. In addition, NetSemantic provides recommendations for mitigating the anomalies, enabling the network to rapidly address the issues and maintain operational stability.

\begin{table*}[t]
\caption{Fault Diagnosis in Different Topological Structures}
\label{5}
\centering
\begin{tabular*}{\textwidth}{@{\extracolsep{\fill}}l *{5}{c} *{5}{c} } 
\toprule[1.2pt]
\multirow{2}{*}{Topology} &\multicolumn{4}{c}{Anomaly Detection} & \multicolumn{4}{c}{Fault Diagnosis} \\ 
\cline{2-5} \cline{6-9}
& Accuracy$ \uparrow$ & Recall$ \uparrow$ & FNR$ \downarrow$ & FPR$ \downarrow $&Accuracy$ \uparrow$ &Recall$ \uparrow$ &FNR$ \downarrow$ &FPR$ \downarrow$ \\
\hline
\hline
\makebox[0.12\textwidth][l]{Star(9-12)}&98.90&98.98&1.02&6.25&92.30& 91.75 & 8.24 & 2.91  \\
\makebox[0.12\textwidth][l]{Ring(9-12)}&98.20&98.27&1.72&5.88&96.90&96.66 &3.39 &1.07\\
\makebox[0.12\textwidth][l]{Mesh(9-12)}&97.04&97.41&2.58&18.57&92.37&93.32&6.67&12.38\\
\makebox[0.12\textwidth][l]{Star(13-16)}&98.40&98.57&1.42&11.11&91.70& 91.11 & 8.89 & 3.01  \\
\makebox[0.12\textwidth][l]{Ring(13-16)}&96.50&96.93&3.06&25.00&91.70&91.01 &8.89 &2.04\\
\makebox[0.12\textwidth][l]{Mesh(13-16)}&95.91&96.27&3.73&26.67&88.74&89.64&10.35&17.89\\
\makebox[0.12\textwidth][l]{Star(17-20)}&92.90&91.16&2.84&23.07&87.00& 86.19 & 13.81 & 5.26  \\
\makebox[0.12\textwidth][l]{Ring(17-20)}&94.80&95.42&4.57&31.17&86.96&86.14 &13.85 &5.26\\
\makebox[0.12\textwidth][l]{Mesh(17-20)}&92.40&92.90&7.09&36.15&84.70&85.91&14.08&25.75\\
\bottomrule[1.2pt]
\end{tabular*}
\end{table*}

\subsection{Ablation Study}
We provide several variants of NetSemantic for ablation analysis:
\begin{itemize}
\item Net-NoKG:NetSemantic without the support of a knowledge graph (GPT-4o), which does not leverage prior knowledge.
\item Net-GPT3.5:NetSemantic based on GPT-3.5 (GPT-3.5).
\item Net-NoSymb:NetSemantic without symbolic validation (GPT-4o).
\item Net:NetSemantic with only semantic descriptions (GPT-4o).
\end{itemize}

\begin{table*}[t]
\caption{Network Type}
\label{6}
\centering
\begin{tabular}{ccccccc}
\toprule[1.2pt]
\textbf{Network}&\textbf{Device Name}&\textbf{ Transmitting Power }&\textbf{ Bandwidth }&\textbf{ Communication protocol }&\textbf{ Range }&\textbf{ Speed }\\
\hline
\hline
Mobile Network&Mobile Phone&23 dBm &20 MHz &LTE &200m&10 m/s\\

Vehicular Ad hoc Networks&Vehicle &30 dBm &10 MHz &802.11p &200m&20 m/s\\

UAV Network&UAV &20 dBm &5 MHz &802.11AC &400m&15 m/s\\

Cellular Network&Base Station&43 dBm &100 MHz &LTE &500m&\diagbox{}{}\\
\bottomrule[1.2pt]
\end{tabular}
\end{table*}

\begin{table*}[t]
\caption{Fault Diagnosis in Different Network Types}
\label{7}
\centering
\begin{tabular*}{\textwidth}{@{\extracolsep{\fill}}l *{5}{c} *{5}{c} } 
\toprule[1.2pt]
\multirow{2}{*}{Network Type} &\multicolumn{4}{c}{Anomaly Detection} & \multicolumn{4}{c}{Fault Diagnosis} \\ 
\cline{2-5} \cline{6-9}
& Accuracy$ \uparrow$ & Recall$ \uparrow$ & FNR$ \downarrow$ & FPR$ \downarrow $&Accuracy$ \uparrow$ &Recall$ \uparrow$ &FNR$ \downarrow$ &FPR$ \downarrow$ \\
\hline
\hline
\makebox[0.12\textwidth][l]{Mobile Network}&90.80&92.97&7.11&28.99&85.41& 87.55 & 13.44 & 34.01  \\
\makebox[0.12\textwidth][l]{Vehicular Ad hoc Networks}&91.49&93.56&1.72&6.44&86.36 &88.11 &11.89&28.89\\
\makebox[0.12\textwidth][l]{UAV Network}&92.70&94.56&5.44&24.00&86.13&87.88&12.11&31.08\\
\makebox[0.12\textwidth][l]{Cellular Network}&94.21&96.11&3.88&6.25& 88.77 & 90.77 & 9.22& 29.98  \\
\bottomrule[1.2pt]
\end{tabular*}
\end{table*}

As shown in Table~\ref{2}, we validate the effectiveness of (1) the NKG (Net-NoNKG) and (2) symbolic validation (Net-NoSymb).

NKG Support: In scenarios where relevant knowledge is lacking, an LLM primarily relies on expert configurations (such as roles, tasks, and steps) to invoke tools and analyze root causes. However, insufficient knowledge reserves can lead to a decline in diagnostic effectiveness. When compared to NetSemantic, which is supported by a knowledge graph, we observed a decrease in diagnostic accuracy by 19.2\% to 64.1\%.
Two key observations emerged: First, without knowledge graph support, there is significantly more redundancy (1.7 times that of NetSemantic). The root cause is that it cannot clearly distinguish between relevant root causes using only context. For example, root causes like "application layer failure" and "data link layer failure" both involve nodes losing information exchange. Correctly identifying them requires specific knowledge about the number and timing of nodes experiencing a loss of information exchange.Second, the absence of a knowledge graph typically results in very general diagnoses (e.g., "nodes experience abnormal information exchange") and fails to accurately identify many anomalies. Furthermore, we found that although LLMs like GPT-4 have been pre-trained on open corpora, they require external knowledge matching (with fine-tuning limited to updating knowledge) to perform specialized tasks such as network fault diagnosis.

Symbolization Support: Compared to NetSemantic without symbolization, NetSemantic with symbolization achieves a diagnostic efficiency improvement of over 13.39\%. This validates the critical role of symbolization in correcting erroneous knowledge matches and identifying device anomalies, significantly enhancing diagnostic accuracy, particularly for cause-related anomalies. During the diagnostic process, independent semantic processing often leads to hallucinations and misjudgments by LLMs. Symbolization can significantly reduce these misjudgments, thereby improving the accuracy of fault diagnosis.
\subsection{Generalization Study}


We evaluated the fault diagnosis performance of NetSemantic across various network topologies and different node scales. As shown in Table~\ref{5}, the topologies included star, ring, and mesh structures, while the node scales ranged from 9-12, 13-16, to 17-20 nodes. The results demonstrated that NetSemantic performed exceptionally well in all types of topologies and node scales. In centralized connection structures, it precisely identified data transmission anomalies between nodes, achieving an impressive diagnostic accuracy of 92.3\%, highlighting its efficiency and reliability in centralized network environments. For ring topologies, where data is transmitted in a unidirectional loop, the diagnostic accuracy reached 91.7\%, effectively capturing communication anomalies even in environments with high interdependency. In highly interconnected and complex mesh network structures, the diagnostic accuracy was 89.49\%. Although the complexity of mesh topologies posed challenges, NetSemantic demonstrated robust anomaly detection capabilities, efficiently extracting abnormal patterns and accurately locating faulty nodes.

The varying complexity of network topologies did have some impact on performance, but NetSemantic consistently delivered diagnostic results within acceptable ranges. Even under conditions of partial connection failures or data redundancy, its exceptional anomaly detection capabilities remained evident. Similarly, as the number of nodes increased, diagnostic complexity also rose; however, NetSemantic maintained its strong adaptability and flexibility, effectively addressing challenges across diverse scales and structures.

Additionally, we conducted extensive tests on the performance of NetSemantic in various network scenarios. As shown in Table~\ref{6}, the network environments include: Mobile Network~\cite{he2025generative}, which consists mainly of mobile devices such as smartphones, tablets, and laptops, using LTE protocol for data transmission; Vehicular Ad hoc Networks (VANETs)~\cite{tang2021comprehensive}, which primarily consists of moving vehicles communicating with each other (V2V) and with infrastructure (V2I) via the 802.11p protocol; UAV Network~\cite{tan2024outage}, consisting mainly of drones and other devices, using the 802.11AC protocol for high-bandwidth data transmission; and Cellular Network~\cite{du2025resource}, which consists of multiple base stations and fixed devices, typically using LTE or 5G technology for data transmission. The scale of the network environment varies from 13 to 16 network devices connected through a mesh structure. Each network environment presents its own unique characteristics and challenges.

Therefore, we conducted an in-depth analysis and compared the performance of NetSemantic across different network environments, as shown in Table~\ref{7}. The experimental results show that NetSemantic performs exceptionally well in all tested environments. In the relatively static Cellular Network environment, the anomaly detection accuracy reached 94.21\%, enabling it to accurately detect potential network faults. In the high-dynamic network environments (Mobile Network, VANETs, UAV Network), NetSemantic achieves a fault diagnosis accuracy of over 85\%, demonstrating its strong adaptability and efficient fault detection capabilities. This ensures the reliability and stability of NetSemantic even in highly dynamic network environments.

\begin{figure}[t]
\centering
\includegraphics[width=0.9\columnwidth]{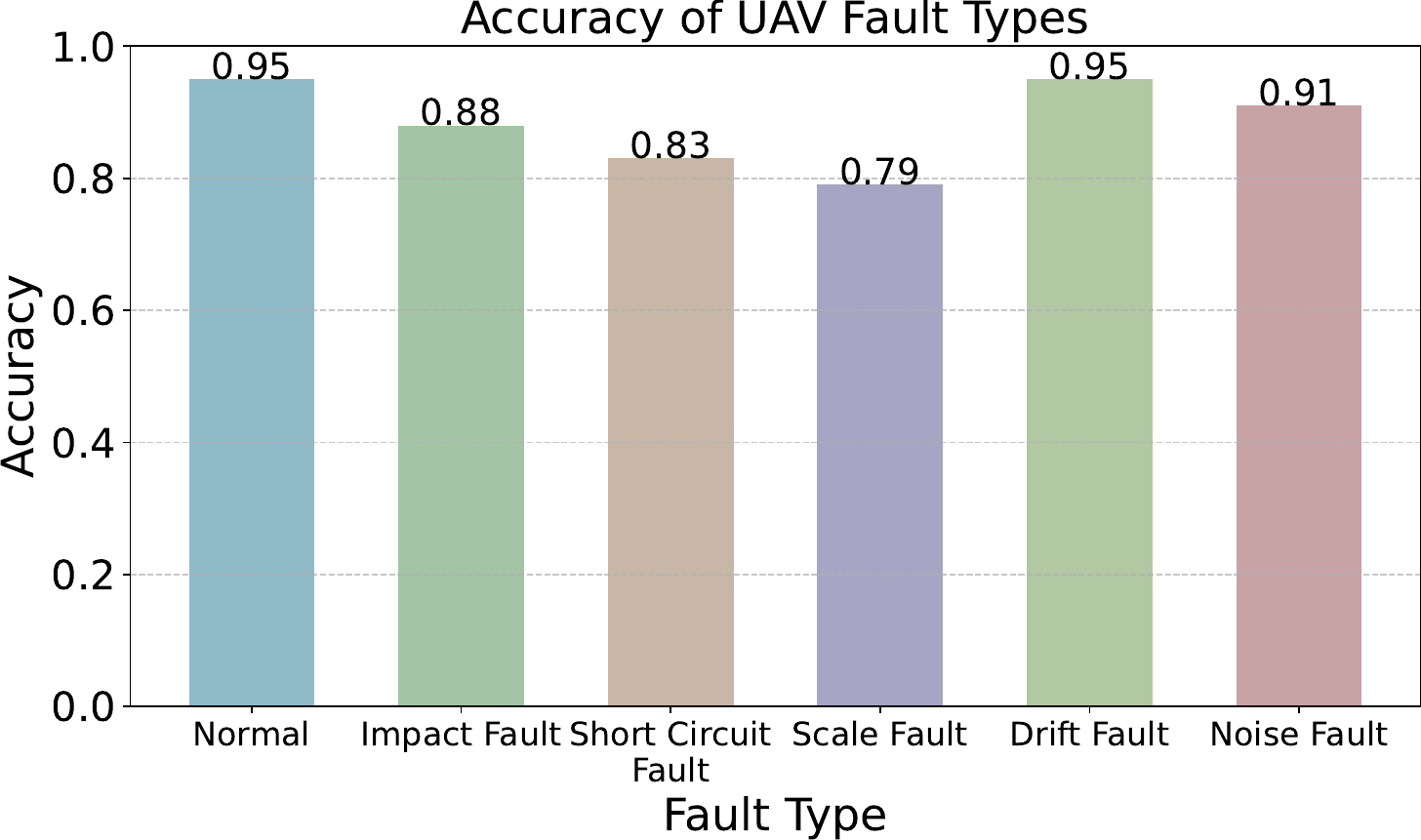}
\caption{UAV Fault Diagnosis.}
\label{fig13}
\end{figure}

To further validate its generalization capability, We conducted additional experiments using a fault dataset from the accelerometer sensors in unmanned aerial vehicle (UAV)\footnote{\url{https://github.com/SmallFlame/uav_faultdata}}. As shown in Fig.~\ref{fig13}, NetSemantic achieved an accuracy of 90\%, showcasing its outstanding fault diagnosis capabilities and potential for flexible anomaly analysis. Beyond generating targeted solutions based on historical data, it also exhibited strong interpretability, allowing it to quickly adapt to new metrics and perspectives.

\section{Analysis and Discussion}


\begin{table*}[htb]
\centering
\caption{Performance on Different Prompts}
\label{9}
\renewcommand{\arraystretch}{1.1} 
\begin{tabularx}{\textwidth}{lXcc}
\toprule[1.2pt]
Prompt      & Template & Anomaly Detection ACC & Fault Diagnosis ACC \\
\hline
\hline
zero-shot   & \textbf{CONTENT} \textless{}task description\textgreater{} \textless{}input\textgreater{}  & 77.42 & 69.36 \\
\hline
general-info & \textbf{INSTRUCTIONS} \textless{}role\textgreater{} \textless{}task description\textgreater{} \textless{}reinforce\textgreater{}\\
            & \textbf{CONTENT} \textless{}input\textgreater{}  \textless{}task confirmation\textgreater{}\\
            & \textbf{CONSTRAINT}  
            \textless{}steps\textgreater{}
            \textless{}rules\textgreater{} & 84.21 & 81.41 \\
\hline
expertise   & \textbf{INSTRUCTIONS} \textless{}role\textgreater{} \textless{}task description\textgreater{} \textless{}reinforce\textgreater{}\\
            & \textbf{CONTENT} \textless{}input\textgreater{}  \textless{}task confirmation\textgreater{}\\
            & \textbf{CONSTRAINT} \textless{}expertise\textgreater{}  
            \textless{}steps\textgreater{}
            \textless{}rules\textgreater{} & 90.20 & 86.41 \\
\hline
NetSemantic & \textbf{INSTRUCTIONS} \textless{}role\textgreater{} \textless{}task description\textgreater{} \textless{}reinforce\textgreater{}\\
            & \textbf{CONTENT} \textless{}input\textgreater{}  \textless{}task confirmation\textgreater{}\\
            & \textbf{CONSTRAINT} \textless{}knowledge\textgreater{}  
            \textless{}steps\textgreater{}
            \textless{}rules\textgreater{}& 96.10 & 89.49 \\
\bottomrule[1.2pt]
\end{tabularx}
\end{table*}

\subsection{Semantic Inquiry}
\begin{figure}[t]
\centering
\includegraphics[width=0.9\columnwidth]{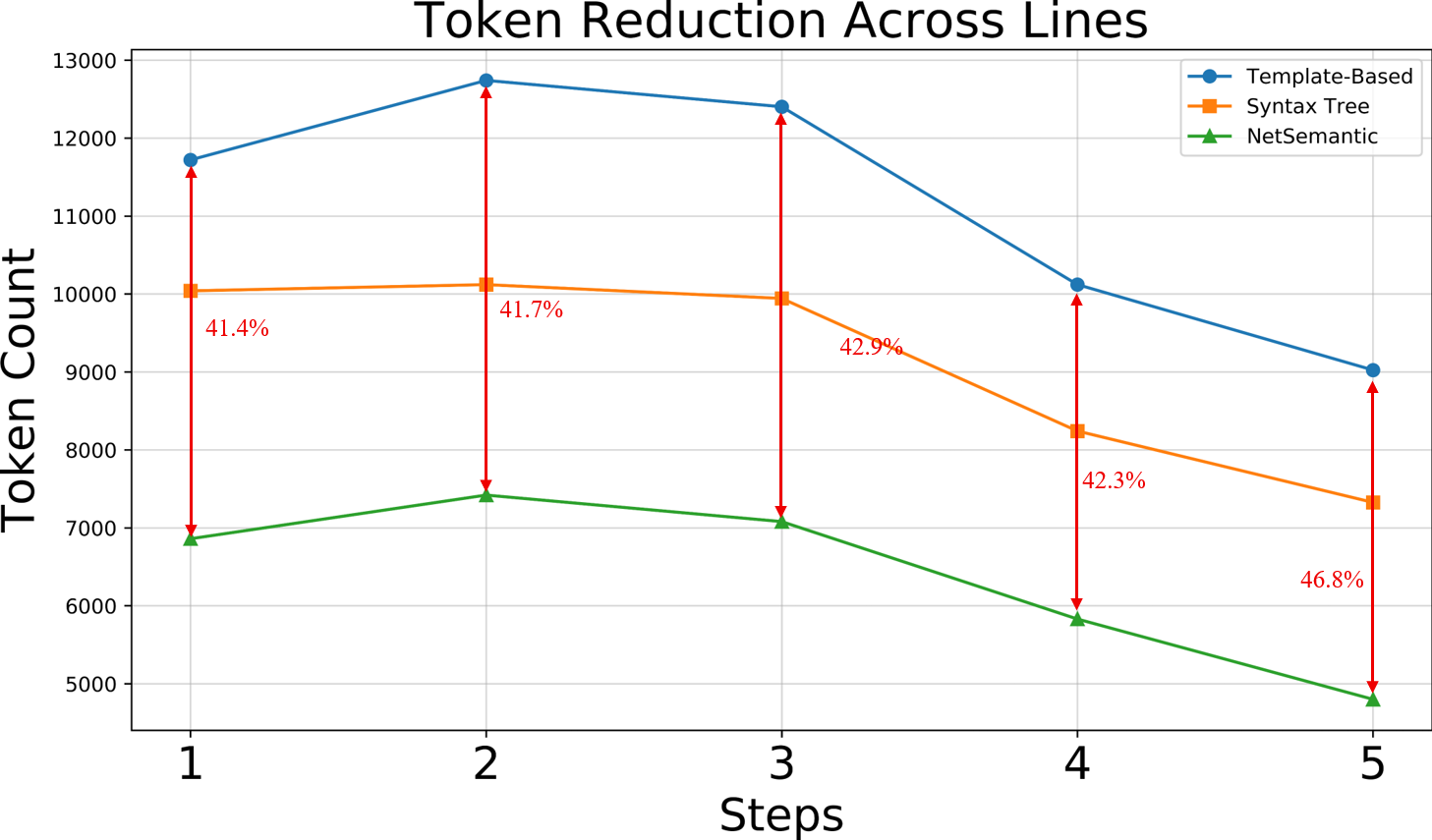}
\caption{Token Input Reduce.}
\label{fig12}
\end{figure}

Prior to this, we proposed two approaches for the process of semanticization:

Template-Based Approach: This method relies on manually designed text templates to semantically transform device data. By mapping feature values, trend information, and specific events in time-series data to a predefined language framework, semanticization is achieved. For example, as illustrated in Fig.~\ref{fig12}, this manual approach is effective for simple data semanticization tasks in specific scenarios. However, subsequent studies revealed significant limitations. The extensibility of manual text templates is poor, making them unsuitable for handling diverse input requirements. Furthermore, as the templates rely heavily on expert knowledge for design, they are costly to maintain and struggle to capture the nuanced features of complex data. Additionally, in complex network environments, template-based descriptions often result in large tokens, leading to resource inefficiencies. Hence, the rigidity of templates restricts their adaptability to complex data patterns and dynamic demands.

Syntax Tree Approach: This method uses structured rules to parse data, decomposing it layer by layer into semantic units and organizing it into text that aligns with natural language logic. For example, we optimized the syntax tree generation rules to improve the expression of various data types. Specifically, we generated three descriptive modes for the latency attribute of nodes, categorizing latency into ranges such as "Normal," "Slight Delay," and "Severe Delay" based on fixed thresholds, and selecting corresponding nodes to generate text based on measured values. While this approach addresses issues such as poor template extensibility and large tokens, our research found limitations. It struggles to capture finer-grained details, and the cost of designing rules and performing syntax parsing increases significantly when dealing with multidimensional or dynamically changing data.

\subsection{The Impact of Prompts and Models}
A common strategy for applying models to specific tasks is model fine-tuning. However, due to its labor-intensive nature and significant resource consumption, this approach is often avoided. As a result, attention has shifted to prompt optimization, which has a considerable impact on the relevance and accuracy of outputs from large language models (LLMs).


As shown in Table~\ref{9}, traditional zero-shot prompting guides the model by directly describing the task and related questions. However, we found that LLMs tend to produce hallucinations in their task descriptions. Additionally, due to a lack of domain-specific knowledge, the accuracy of these descriptions often leaves room for improvement. Based on the provided content, the prompts can be categorized as either expert prompts (which incorporate specialized knowledge) or general-information prompts (which do not). Although these methods enhance diagnostic accuracy to a certain extent, the rapid pace of network information updates renders expert knowledge outdated. Consequently, they still face significant challenges in dynamic network environments.

In this context, we turned to explore leveraging the knowledge summarized by the LLM itself to unlock its potential. We refer to this as a self-inspired prompt template, which also operates in a zero-shot manner. Specifically, since network environments are complex and dynamic, with new devices being added, devices being upgraded, and configuration information being updated at any time, we use the LLM to continually update its information. This knowledge is then summarized and stored in a NKG. For example, when new latency-sensitive devices are introduced into the network, we use the LLM to analyze their attributes and derive performance metrics corresponding to various parameter values.

We also explored the impact of different foundational models. The NetSemantic model based on GPT-4 outperformed its GPT-3.5 counterpart, achieving ACC scores of 96.1\% and 89.49\%, compared to 83.31\% and 70.42\%, respectively. These results demonstrate that GPT-4 excels in data collection and reasoning capabilities across various sources, such as multimodal metrics, when compared to GPT-3.5.

\subsection{Potential Applications of Network Semanticization and LLMs}
Beyond network fault diagnosis, network semanticization combined with LLMs presents promising opportunities in various networking domains. By structuring network data into meaningful semantic representations and leveraging LLMs’ advanced reasoning and text generation capabilities, these approaches can enhance network performance prediction, traffic engineering, and the construction of digital twin networks.

One promising direction is network performance prediction, where semanticization enables a structured understanding of key performance indicators such as latency and throughput. LLMs further enhance this process by reasoning over semantic representations, identifying deep patterns in network behavior, and generating predictive insights. By integrating historical and real-time data, these models can improve forecasting accuracy and support proactive network adjustments to enhance stability and efficiency.

Another area of exploration is traffic engineering, where network semanticization supports dynamic optimization by interpreting real-time traffic flows and predicting congestion. LLMs enhance this capability by analyzing traffic semantics, generating adaptive policies, and automating decision-making processes. Through semantic analysis and natural language reasoning, LLMs enable more intelligent routing adjustments, bandwidth allocation, and Quality of Service (QoS) optimizations, making traffic management more responsive and adaptable to network changes.

A further compelling application lies in digital twin networks, where network semanticization plays a crucial role in constructing digital replicas of physical networks. LLMs contribute by interpreting and generating detailed semantic models, enabling more precise network state descriptions and simulations. Their ability to generate structured network insights allows for improved real-time monitoring, anomaly detection, and predictive analysis. As a result, these semantically driven digital twins enhance network adaptability and operational efficiency, facilitating more accurate and scalable network management.

\section{Conclusion}
In this paper, we proposed NetSemantic, a plug-and-play intelligent network information semantic fault diagnosis method based on large language models, which is independent of data dependencies. For the first time, it unifies multimodal network data into a semantic and symbolic representations, enabling zero-shot fault diagnosis. Experimental results demonstrate that NetSemantic achieves significant improvements over baseline methods and other fault diagnosis systems.

\bibliographystyle{IEEEtran}
\bibliography{reference}

\begin{thebibliography}{10}
\providecommand{\url}[1]{#1}
\csname url@samestyle\endcsname
\providecommand{\newblock}{\relax}
\providecommand{\bibinfo}[2]{#2}
\providecommand{\BIBentrySTDinterwordspacing}{\spaceskip=0pt\relax}
\providecommand{\BIBentryALTinterwordstretchfactor}{4}
\providecommand{\BIBentryALTinterwordspacing}{\spaceskip=\fontdimen2\font plus
\BIBentryALTinterwordstretchfactor\fontdimen3\font minus \fontdimen4\font\relax}
\providecommand{\BIBforeignlanguage}[2]{{%
\expandafter\ifx\csname l@#1\endcsname\relax
\typeout{** WARNING: IEEEtran.bst: No hyphenation pattern has been}%
\typeout{** loaded for the language `#1'. Using the pattern for}%
\typeout{** the default language instead.}%
\else
\language=\csname l@#1\endcsname
\fi
#2}}
\providecommand{\BIBdecl}{\relax}
\BIBdecl

\bibitem{li2020research}
M.~Li, M.~Huo, X.~Cheng, and L.~Xu, ``Research and application of ai in 5g network operation and maintenance,'' in \emph{2020 IEEE Intl Conf on Parallel \& Distributed Processing with Applications, Big Data \& Cloud Computing, Sustainable Computing \& Communications, Social Computing \& Networking (ISPA/BDCloud/SocialCom/SustainCom)}.\hskip 1em plus 0.5em minus 0.4em\relax IEEE, 2020, pp. 1420--1425.

\bibitem{notaro2021survey}
P.~Notaro, J.~Cardoso, and M.~Gerndt, ``A survey of aiops methods for failure management,'' \emph{ACM Transactions on Intelligent Systems and Technology (TIST)}, vol.~12, no.~6, pp. 1--45, 2021.

\bibitem{qureshi2023toward}
H.~N. Qureshi, U.~Masood, M.~Manalastas, S.~M.~A. Zaidi, H.~Farooq, J.~Forgeat, M.~Bouton, S.~Bothe, P.~Karlsson, A.~Rizwan \emph{et~al.}, ``Toward addressing training data scarcity challenge in emerging radio access networks: A survey and framework,'' \emph{IEEE Communications Surveys \& Tutorials}, vol.~25, no.~3, pp. 1954--1990, 2023.

\bibitem{tu2024towards}
T.~Tu, S.~Azizi, D.~Driess, M.~Schaekermann, M.~Amin, P.-C. Chang, A.~Carroll, C.~Lau, R.~Tanno, I.~Ktena \emph{et~al.}, ``Towards generalist biomedical ai,'' \emph{Nejm Ai}, vol.~1, no.~3, p. AIoa2300138, 2024.

\bibitem{zhou2024lawgpt}
Z.~Zhou, J.-X. Shi, P.-X. Song, X.-W. Yang, Y.-X. Jin, L.-Z. Guo, and Y.-F. Li, ``Lawgpt: A chinese legal knowledge-enhanced large language model,'' \emph{arXiv preprint arXiv:2406.04614}, 2024.

\bibitem{wang2024chattime}
C.~Wang, Q.~Qi, J.~Wang, H.~Sun, Z.~Zhuang, J.~Wu, L.~Zhang, and J.~Liao, ``Chattime: A unified multimodal time series foundation model bridging numerical and textual data,'' \emph{arXiv preprint arXiv:2412.11376}, 2024.

\bibitem{hu2025context}
Y.~Hu, Q.~Li, D.~Zhang, J.~Yan, and Y.~Chen, ``Context-alignment: Activating and enhancing llm capabilities in time series,'' \emph{arXiv preprint arXiv:2501.03747}, 2025.

\bibitem{jin2023time}
M.~Jin, S.~Wang, L.~Ma, Z.~Chu, J.~Y. Zhang, X.~Shi, P.-Y. Chen, Y.~Liang, Y.-F. Li, S.~Pan \emph{et~al.}, ``Time-llm: Time series forecasting by reprogramming large language models,'' \emph{arXiv preprint arXiv:2310.01728}, 2023.

\bibitem{tang2024semi}
F.~Tang, L.~Luo, Z.~Guo, Y.~Li, M.~Zhao, and N.~Kato, ``Semi-distributed network fault diagnosis based on digital twin network in highly dynamic heterogeneous networks,'' \emph{IEEE Transactions on Mobile Computing}, 2024.

\bibitem{tang2024large}
F.~Tang, X.~Wang, X.~Yuan, L.~Luo, M.~Zhao, and N.~Kato, ``Large language model (llm) assisted end-to-end network health management based on multi-scale semanticization,'' \emph{arXiv preprint arXiv:2406.08305}, 2024.

\bibitem{wu2024netllm}
D.~Wu, X.~Wang, Y.~Qiao, Z.~Wang, J.~Jiang, S.~Cui, and F.~Wang, ``Netllm: Adapting large language models for networking,'' in \emph{Proceedings of the ACM SIGCOMM 2024 Conference}, 2024, pp. 661--678.

\bibitem{chen2024automatic}
Y.~Chen, H.~Xie, M.~Ma, Y.~Kang, X.~Gao, L.~Shi, Y.~Cao, X.~Gao, H.~Fan, M.~Wen \emph{et~al.}, ``Automatic root cause analysis via large language models for cloud incidents,'' in \emph{Proceedings of the Nineteenth European Conference on Computer Systems}, 2024, pp. 674--688.

\bibitem{jia2024gpt4mts}
F.~Jia, K.~Wang, Y.~Zheng, D.~Cao, and Y.~Liu, ``Gpt4mts: Prompt-based large language model for multimodal time-series forecasting,'' in \emph{Proceedings of the AAAI Conference on Artificial Intelligence}, vol.~38, no.~21, 2024, pp. 23\,343--23\,351.

\bibitem{yedavalli2011application}
R.~K. Yedavalli and R.~K. Belapurkar, ``Application of wireless sensor networks to aircraft control and health management systems,'' \emph{Journal of Control Theory and Applications}, vol.~9, pp. 28--33, 2011.

\bibitem{lee2006network}
W.~L. Lee, A.~Datta, R.~Cardell-Oliver \emph{et~al.}, ``Network management in wireless sensor networks,'' \emph{Handbook of Mobile Ad Hoc and Pervasive Communications}, pp. 1--20, 2006.

\bibitem{tariq2019detecting}
S.~Tariq, S.~Lee, Y.~Shin, M.~S. Lee, O.~Jung, D.~Chung, and S.~S. Woo, ``Detecting anomalies in space using multivariate convolutional lstm with mixtures of probabilistic pca,'' in \emph{Proceedings of the 25th ACM SIGKDD international conference on knowledge discovery \& data mining}, 2019, pp. 2123--2133.

\bibitem{ren2019time}
H.~Ren, B.~Xu, Y.~Wang, C.~Yi, C.~Huang, X.~Kou, T.~Xing, M.~Yang, J.~Tong, and Q.~Zhang, ``Time-series anomaly detection service at microsoft,'' in \emph{Proceedings of the 25th ACM SIGKDD international conference on knowledge discovery \& data mining}, 2019, pp. 3009--3017.

\bibitem{lei2020applications}
Y.~Lei, B.~Yang, X.~Jiang, F.~Jia, N.~Li, and A.~K. Nandi, ``Applications of machine learning to machine fault diagnosis: A review and roadmap,'' \emph{Mechanical systems and signal processing}, vol. 138, p. 106587, 2020.

\bibitem{wardat2021deeplocalize}
M.~Wardat, W.~Le, and H.~Rajan, ``Deeplocalize: Fault localization for deep neural networks,'' in \emph{2021 IEEE/ACM 43rd International Conference on Software Engineering (ICSE)}.\hskip 1em plus 0.5em minus 0.4em\relax IEEE, 2021, pp. 251--262.

\bibitem{jeong2023anomalybert}
Y.~Jeong, E.~Yang, J.~H. Ryu, I.~Park, and M.~Kang, ``Anomalybert: Self-supervised transformer for time series anomaly detection using data degradation scheme,'' \emph{CoRR}, vol. abs/2305.04468, 2023.

\bibitem{cao2024advanced}
K.~Cao, T.~Zhang, and J.~Huang, ``Advanced hybrid lstm-transformer architecture for real-time multi-task prediction in engineering systems,'' \emph{Scientific Reports}, vol.~14, no.~1, p. 4890, 2024.

\bibitem{xue2023promptcast}
H.~Xue and F.~D. Salim, ``Promptcast: A new prompt-based learning paradigm for time series forecasting,'' \emph{IEEE Transactions on Knowledge and Data Engineering}, vol.~36, no.~11, pp. 6851--6864, 2023.

\bibitem{liu2024can}
L.~Liu, S.~Yu, R.~Wang, Z.~Ma, and Y.~Shen, ``How can large language models understand spatial-temporal data?'' \emph{arXiv preprint arXiv:2401.14192}, 2024.

\bibitem{reimers2019sentence}
N.~Reimers and I.~Gurevych, ``Sentence-bert: Sentence embeddings using siamese bert-networks,'' \emph{arXiv preprint arXiv:1908.10084}, 2019.

\bibitem{gavrilo2024using}
D.~Gavrilo, C.~Johansson, T.~Petrovich, G.~Martins, and S.~Edwards, ``Using dynamic token embedding compression to optimize inference process in large language models,'' 2024.

\bibitem{de2008z3}
L.~De~Moura and N.~Bj{\o}rner, ``Z3: An efficient smt solver,'' in \emph{International conference on Tools and Algorithms for the Construction and Analysis of Systems}.\hskip 1em plus 0.5em minus 0.4em\relax Springer, 2008, pp. 337--340.

\bibitem{jatnika2019word2vec}
D.~Jatnika, M.~A. Bijaksana, and A.~A. Suryani, ``Word2vec model analysis for semantic similarities in english words,'' \emph{Procedia Computer Science}, vol. 157, pp. 160--167, 2019.

\bibitem{islam2024gpt}
R.~Islam and O.~M. Moushi, ``Gpt-4o: The cutting-edge advancement in multimodal llm,'' \emph{Authorea Preprints}, 2024.

\bibitem{ouyang2022training}
L.~Ouyang, J.~Wu, X.~Jiang, D.~Almeida, C.~Wainwright, P.~Mishkin, C.~Zhang, S.~Agarwal, K.~Slama, A.~Ray \emph{et~al.}, ``Training language models to follow instructions with human feedback,'' \emph{Advances in neural information processing systems}, vol.~35, pp. 27\,730--27\,744, 2022.

\bibitem{he2025generative}
L.~He, G.~Sun, D.~Niyato, H.~Du, F.~Mei, J.~Kang, M.~Debbah, and Z.~Han, ``Generative ai for game theory-based mobile networking,'' \emph{IEEE Wireless Communications}, vol.~32, no.~1, pp. 122--130, 2025.

\bibitem{tang2021comprehensive}
F.~Tang, B.~Mao, N.~Kato, and G.~Gui, ``Comprehensive survey on machine learning in vehicular network: Technology, applications and challenges,'' \emph{IEEE Communications Surveys \& Tutorials}, vol.~23, no.~3, pp. 2027--2057, 2021.

\bibitem{tan2024outage}
J.~Tan, F.~Tang, M.~Zhao, and N.~Kato, ``Outage probability, performance, fairness analysis of space-air-ground integrated network (sagin): Uav altitude and position angle,'' \emph{IEEE Transactions on Wireless Communications}, 2024.

\bibitem{du2025resource}
Y.~Du, L.~Yang, and Y.~Luo, ``Resource allocation based on optimized cellular network ap layout for visible light communication heterogeneous network,'' \emph{The Journal of Supercomputing}, vol.~81, no.~1, pp. 1--23, 2025.

\end{thebibliography}

\end{document}